\documentclass[aps,floats]{revtex4}
\usepackage{amsmath,amssymb}
\usepackage{graphicx,epsfig}

\begin{document}
\bibliographystyle {plain}

\def\oppropto{\mathop{\propto}} 
\def\opsimeq{\mathop{\simeq}}
\def\opoverderline{\mathop{\overline}}
\def\operarrow{\mathop{\longrightarrow}}
\def\opsim{\mathop{\sim}}

\def\fig#1#2{\includegraphics[height=#1]{#2}}
\def\figx#1#2{\includegraphics[width=#1]{#2}}


\title{ One-dimensional Ising spin-glass with power-law interaction : \\
 real-space renormalization at zero temperature
  } 


 \author{ C\'ecile Monthus }
  \affiliation{Institut de Physique Th\'{e}orique,\\
  CNRS and CEA Saclay \\
 91191 Gif-sur-Yvette, France}

\begin{abstract}
For the one-dimensional long-ranged Ising spin-glass with random couplings decaying with the distance $r$ as $J(r) \sim r^{-\sigma}$ and distributed with the L\'evy symmetric stable distribution of index $1 <\mu \leq 2$ (including the usual Gaussian case $\mu=2$), we consider the region $\sigma>1/\mu$ where the energy is extensive.  We study two real space renormalization procedures at zero temperature, namely a simple box decimation that leads to explicit calculations, and a strong disorder decimation that can be studied numerically on large sizes. The droplet exponent governing the scaling of the renormalized couplings $J_L \propto L^{\theta_{\mu}(\sigma)}$ is found to be $\theta_{\mu}(\sigma)=\frac{2}{\mu}-\sigma$ whenever the long-ranged couplings are relevant $\theta_{\mu}(\sigma) \geq -1$. For the statistics of the ground state energy $E_L^{GS}$ over disordered samples, we obtain that the droplet exponent $\theta_{\mu}(\sigma) $ governs the leading correction to extensivity of the averaged value $\overline{E_L^{GS}} \simeq L e_0 +L^{\theta_{\mu}(\sigma)} e_1$. The characteristic scale of the fluctuations around this average is of order $L^{\frac{1}{\mu}}$, and the rescaled variable $u=(E_L^{GS}-\overline{E_L^{GS}})/L^{\frac{1}{\mu}}$ is Gaussian distributed for $\mu=2$, or displays the negative power-law tail in $1/(-u)^{1+\mu}$ for $u \to -\infty$ in the L\'evy case $1<\mu<2$. Finally we apply the zero-temperature renormalization procedure to the related Dyson hierarchical spin-glass model where the same droplet exponent appears.

\end{abstract}

\maketitle

\section{ Introduction}

In the field of classical spin-glasses \cite{binder_young,young,newman_stein_book}, 
real space renormalization
procedures have been much studied in the Migdal-Kadanoff approximation 
\cite{southern_young,young_St,mckay,BM_MK,Gardnersg,banavar,bokil,muriel,thill,ritortMK,boettcher_mk,us_tail,jorg,us_sgferro} where hypercubic lattices are effectively replaced by hierarchical
fractal lattices whose structure is by construction exactly renormalizable
\cite{MKRG,berker,hierarchical}. But if one insists on keeping the
hypercubic lattice in dimension $d>1$, the precise definition of an
appropriate renormalization procedure has remained very difficult.
This is all the more annoying when taking into account that 
the droplet scaling theory of spin-glasses \cite{mcmillan,bray_moore,fisher_huse}
is based on renormalization scaling ideas, and in particular on the 
droplet exponent $\theta$ that governs the scaling 
of the renormalized random coupling $J_L$ with the length $L$
 \begin{eqnarray}
J_L \propto L^{\theta}
\label{deftheta}
\end{eqnarray}

Since the whole low temperature spin-glass phase is governed by the
zero-temperature fixed point characterized by the droplet exponent of Eq. \ref{deftheta},
it seems clear that 
the first goal should be to define an appropriate explicit renormalization procedure
directly at zero temperature and to test the value of the droplet exponent $\theta$ obtained.
For short-ranged models on hypercubic lattices in dimension $d$, the droplet
exponent $\theta^{SR}(d)$ is known only numerically for $d > 1$
(see for instance \cite{boettcher} and references therein)
and is well below the simple upper-bound \cite{fisher_huse}
 \begin{eqnarray}
\theta^{SR}(d) \leq \frac{d-1}{2}
\label{boundtheta}
\end{eqnarray}
Note that even in $d=1$, the exactly known droplet exponent \cite{bray_moore}
 \begin{eqnarray}
\theta^{SR}(d=1) =-1
\label{thetaSRd1}
\end{eqnarray}
is well below the bound of Eq. \ref{boundtheta}.
The situation is better for long-ranged spin-glasses when the 
variance of the initial couplings
decays only as a power-law with respect to the distance 
\begin{eqnarray}
\overline{J^2_{i,j}} \propto \frac{1}{\vert j-i \vert^{2 \sigma}}
\label{variancesigma}
\end{eqnarray}
where the exponent $\sigma$ has to satisfy the bound
\begin{eqnarray}
\sigma > \frac{d}{2}
\label{regionsigmaextensive}
\end{eqnarray}
in order to ensure that the energy is extensive in the number of spins.
The corresponding droplet exponent $\theta^{LR}(d,\sigma) $ is then believed to be known exactly 
\cite{fisher_huse,BMY} in the region of parameters where it is bigger than the
corresponding short-ranged droplet exponent $\theta^{SR}(d)$ 
 \begin{eqnarray}
\theta^{LR}(d,\sigma) = d-\sigma \ \ \ \ \ \  {\rm if } \ \ \theta^{LR}(d,\sigma)>\theta^{SR}(d)
\label{thetaLR}
\end{eqnarray}
In particular in dimension $d=1$ where the short-ranged droplet exponent $\theta^{SR}(d=1)$
is given by Eq. \ref{thetaSRd1}, the long-ranged droplet exponent is then exactly
known in the whole extensive region $\sigma>\frac{1}{2}$ (Eq. \ref{regionsigmaextensive})
by
\begin{eqnarray}
\theta^{LR}(d=1,\sigma) && = 1-\sigma \ \ \ \ \ \ \ \ \ \ \ \ \ \ \ \ \ \ {\rm for } \ \ \frac{1}{2} < \sigma < 2
\nonumber \\
\theta^{LR}(d=1,\sigma) && = \theta^{SR}(d=1)=-1 \ \  {\rm for } \ \  2 \leq \sigma  
\label{thetaLRd1}
\end{eqnarray}

The goal of the present paper is to study explicit zero-temperature
renormalization procedures for the long-ranged one-dimensional spin-glass
and to see whether they are able to reproduce the results of Eq. \ref{thetaLRd1}.
The paper is organized as follows.
In section \ref{sec_model}, the one-dimensional long-ranged spin-glass model is described,
both for the case of finite variance (Eq. \ref{variancesigma}) and for the case 
of L\'evy distribution with infinite variance.
In section \ref{sec_box}, the simplest block decimation is shown to reproduce
Eq. \ref{thetaLRd1} only in the region of positive droplet exponent.
In section \ref{sec_strong}, we introduce a strong disorder decimation which is able to reproduce
Eq. \ref{thetaLRd1} even in the region of negative droplet exponent.
 In section \ref{sec_magnetic}, we discuss the stability with respect to an external magnetic field.
In section \ref{sec_energy}, we analyze the statistics of the ground state energy over samples. Finally in section \ref{sec_dyson}, we apply the zero-temperature
renormalization procedure to the related Dyson hierarchical spin-glass model.
Our conclusions are summarized in section \ref{sec_conclusion}.

\section{ Model and notations  }

\label{sec_model}

In this paper, we consider the one-dimensional model of classical spins $S_i=\pm 1$
defined by the energy
\begin{eqnarray}
 E  = - \sum_{-\infty \leq n <m \leq +\infty} J_{n m} S_n S_m 
\label{HSGini}
\end{eqnarray}
The random couplings decay as a power-law of the distance with exponent $\sigma$
\begin{eqnarray}
J_{n m} =  \frac{\epsilon_{n m}}{(m-n)^{\sigma}}
\label{defjijini}
\end{eqnarray}
and the amplitudes
 $\epsilon_{n m} $ are independent identical random variables of zero mean.

\subsection { Gaussian distribution }

The case where the amplitudes $\epsilon_{n m} $ of Eq. \ref{defjijini} have a Gaussian distribution
 \begin{eqnarray}
L_{2}(\epsilon)  =  \frac{1}{\sqrt{4 \pi} } e^{- \frac{\epsilon^2}{4}}
\label{gaussian}
\end{eqnarray}
and thus a finite variance (Eq. \ref{variancesigma})
has been much studied in the literature \cite{kotliar,leuzzi99,KY,KYgeom,KKLH,KKLJH,Katz,leuzzietal,KYalmeida,Yalmeida,KDYalmeida,LRmoore,KHY,KH,mori,wittmann,us_overlaptyp,us_dynamic,us_chaos}.
As recalled in the Introduction, the condition for the extensivity of the energy is given by Eq.
\ref{regionsigmaextensive}
\begin{eqnarray}
\sigma > \frac{1}{2}
\label{regionsigmaextensivegauss}
\end{eqnarray}
 and the droplet exponent is expected to be exactly
given by Eq. \ref{thetaLR}. Note however that the numerical measures
 via Monte-Carlo on sizes $L \leq 256$
(see Fig. 13 and Table III of \cite{KY}) are not a clear support of this theoretical expectation,
in particular in the region $\sigma \to (1/2)^+$ where the theoretical prediction of Eq. \ref{thetaLR} corresponds to $\theta^{LR}(d=1,\sigma\to (1/2)^+) \to (1/2)^-$, whereas the numerical results of \cite{KY} display a saturation around $\theta \simeq 0.3$. The origin of this discrepancy has remained unclear over the years.
The interpretation proposed in  \cite{KY} is that
Eq. \ref{thetaLRd1} is nevertheless exact in the whole region 
$\frac{1}{2} < \sigma < 2 $
as predicted by the theoretical derivations \cite{fisher_huse,BMY},
and despite their numerical results \cite{KY} .
Another interpretation could be that the saturation seen in the numerics is
meaningful, and that Eq. \ref{thetaLRd1} is not exact in the whole region 
$\frac{1}{2} < \sigma < 2 $,
but we are not aware of any such statement in the literature.

\subsection { L\'evy distribution with infinite variance }
 
In this paper, we will also consider the case where the amplitudes $\epsilon_{n m} $
of Eq. \ref{defjijini} are drawn with the L\'evy 
symmetric stable law $L_{\mu}(\epsilon) $ of index $0<\mu \leq 2$
defined by its Fourier Transform ${\hat L}_{\mu} (k)$
 \begin{eqnarray}
{\hat L}_{\mu} (k) && \equiv \overline{ e^{i k \epsilon} } =  \int_{-\infty}^{+\infty} d \epsilon e^{i k \epsilon} L_{\mu}(\epsilon) = e^{ - \vert k \vert^{\mu} }
\nonumber \\
L_{\mu}(\epsilon) && = \int_{-\infty}^{+\infty} \frac{dk}{2 \pi} e^{-i k \epsilon  - \vert k \vert^{\mu} }
\label{levy}
\end{eqnarray}
The case $\mu=2$ of course corresponds to the Gaussian distribution of Eq. \ref{gaussian},
whereas the other cases $0<\mu<2$ correspond to distributions with the following power-law tail
 \begin{eqnarray}
L_{\mu<2}(\epsilon) && \opsimeq_{ \epsilon  \to \pm \infty} \frac{A_{\mu}}{\vert \epsilon \vert^{1+\mu}}
\nonumber \\
A_{\mu} && \equiv \frac{\Gamma(1+\mu)}{\pi} \sin \left(\frac{\pi \mu}{2} \right)
\label{levytail}
\end{eqnarray}
The case of L\'evy distributions of couplings has been already studied for 
the mean-field fully connected geometry \cite{cizeau,janzen,mezard,neri,boettcher_levy}
and for the nearest-neighbor model in dimension $d=3$ \cite{andresen}.

For further purposes, it is convenient to introduce the following notation
for the characteristic scale $\Delta(r)$ of the initial model as a function of the distance $r$
 \begin{eqnarray}
\Delta(r)= \frac{1}{r^{\sigma}}
\label{deltar}
\end{eqnarray}
Then the coupling $J_{n,n+r} $ between two sites separated by the distance $r$
is distributed with the L\'evy stable law of scale $\Delta(r) $ and of index $\mu$
 \begin{eqnarray}
P_r(J_{n,n+r}) && = \frac{1}{\Delta(r)} L_{\mu} \left( \frac{J_{i,i+r}}{\Delta(r)} \right)
\label{levyj}
\end{eqnarray}
so that its Fourier transform reads
 \begin{eqnarray}
{\hat P}_r (k) && \equiv \overline{ e^{i k J_{n,n+r} } }  = e^{ - \vert k \Delta(r) \vert^{\mu} }
\label{fourierj}
\end{eqnarray}

\subsection { Condition to ensure the extensivity of the energy }

To have an extensive energy (Eq. \ref{HSGini})
with respect to the number of spins,
one should first impose that the local field $h_n$ seen by a given spin $S_n$
\begin{eqnarray}
h_n =  \sum_{ m \ne n} J_{n m}  S_m 
\label{hloc}
\end{eqnarray}
 remains finite in the thermodynamic limit.
To evaluate the scaling of $h_n$, one can assume that the value of $S_m$ is a random sign 
independent of $ J_{n m}$ so that the Fourier transform of its distribution simply reads
(Eq \ref{fourierj})
\begin{eqnarray}
\overline{ e^{i k h_n} }  \simeq \prod_{ m \ne n} \overline{ e^{i k J_{n m}} } =
 \prod_{ m \ne n} e^{-  \vert \frac{k}{\vert m-n\vert^{ \sigma}} \vert^{\mu} }
=  e^{-\vert k \vert^{\mu} \sum_{m \ne n}  \frac{ 1 }{\vert m-n\vert ^{ \sigma \mu}}}
\label{hlocfourier}
\end{eqnarray}
The sum in the exponential is convergent for 
\begin{eqnarray}
\sigma \mu > 1
\label{sigmahloc}
\end{eqnarray}
Then the ground state energy $E_N$ for a system of $N$ spins
will scale as the sum over $N$ absolute values of the local fields
\begin{eqnarray}
E_N \simeq - \sum_{n=1}^N \vert h_n \vert
\label{sumhloc}
\end{eqnarray}
For $\mu<2$, the distribution of the local field of Eq. \ref{hlocfourier}
displays the same power-law tail of Eq. \ref{levytail}
 \begin{eqnarray}
P_{\mu<2}(h) && \oppropto_{ \vert h \vert  \to  \infty} \frac{1}{\vert h \vert^{1+\mu}}
\label{levytailh}
\end{eqnarray}
so that we have to distinguish two cases :

(i) for $1 < \mu < 2$, the averaged value of $\vert h_n \vert$ is finite,
so that the energy of Eq. \ref{sumhloc} is indeed extensive 
whenever the condition of Eq. \ref{sigmahloc} is satisfied.

(ii) for $0< \mu \leq 1$, the averaged value of $\vert h_n \vert$ is infinite,
so that the energy of Eq. \ref{sumhloc} is not extensive, but grows 
more rapidly as $N^{\frac{1}{\mu}}$ for $\mu<1$ and as $ (N \ln N)$ for $\mu=1$.

In summary for $0 < \mu \leq 1$, the long-ranged L\'evy spin-glass of Eq. \ref{HSGini}
cannot be made extensive, and will not be considered anymore in the following.
In this paper we will focus only the region
\begin{eqnarray}
&& 1 < \mu \leq 2
\nonumber \\
&& \sigma  > \frac{1}{\mu}
\label{regionmusigma}
\end{eqnarray}
 where the long-ranged L\'evy spin-glass of Eq. \ref{HSGini} has an extensive energy.
For the Gaussian case $\mu=2$ (Eq \ref{gaussian}), the condition of Eq. \ref{regionmusigma}
 corresponds to the known boundary of the extensive region (Eq. \ref{regionsigmaextensivegauss}) as it should.

\section{ Block Renormalization at zero temperature }

\label{sec_box}

\subsection{ Decimation rule using blocks of size $b=2$  }

We consider blocks of size $b=2$ containing two neighboring spins $(S_{2n-1},S_{2n})$.
The corresponding internal energy of this block
\begin{eqnarray}
E^{int}_{2n-1,2n} = -J_{2n-1,2n} S_{2n-1} S_{2n}
\label{eint}
\end{eqnarray}
can be minimized by the choice
\begin{eqnarray}
S_{2n-1} =  S_{2n} {\rm sign } (J_{2n-1,2n})
\label{decimb2}
\end{eqnarray}
Eliminating all the odd spins with this rule, the total energy of Eq. \ref{HSGini}
becomes
\begin{eqnarray}
 E  = - \sum_n \vert J_{2n-1,2n} \vert - \sum_{-\infty \leq n <m \leq +\infty} J^{(1)}_{2n, 2m} S_{2n} S_{2n} 
\label{HSG1pas}
\end{eqnarray}
with the following renormalized couplings between the remaining even spins
\begin{eqnarray}
J_{2n,2 m}^{(1)} && = J_{2n,2 m} + {\rm sgn} (J_{2n-1, 2n}){\rm sgn} (J_{2m-1, 2m}) J_{2n-1,2 m-1}
\nonumber \\
&&  +{\rm sgn} (J_{2n-1, 2n}) J_{2n-1,2 m}
+ {\rm sgn} (J_{2m-1, 2m}) J_{2n,2 m-1}
\label{rgjfirst}
\end{eqnarray}
Since all the initial couplings are of random signs and statistically independent,
one obtains that the renormalized couplings are also statistically independent.
Moreover the distribution of the renormalized coupling $J_{2n,2 m}^{(1)}$
can be obtained via its Fourier Transform (Eq. \ref{fourierj})
\begin{eqnarray}
\overline{ e^{i k J_{2n,2 m}^{(1)}}} && =\left(  \overline{  e^{i k J_{2n,2 m} } } \right)
\left( \overline{ e^{i k  J_{2n-1,2 m-1} } } \right)
\left(\overline{ e^{i k  J_{2n-1,2 m} }} \right)
\left( \overline{ e^{i k  J_{2n,2 m-1} }} \right)
\nonumber \\ && = e^{ - \vert k \vert^{\mu} \left[ \frac{2}{(2m-2n)^{\mu \sigma}}
+ \frac{1}{(2m-2n+1)^{\mu \sigma}} + \frac{1}{(2m-2n-1)^{\mu \sigma}} \right] }
\label{fourierrgjfirst}
\end{eqnarray}
So the renormalized coupling $J_{2n,2 n+2r}^{(1)}$ between two even sites separated
by a distance $(2r)$ is distributed with the same L\'evy stable law $L_{\mu}$
of the initial coupling, but with the renormalized characteristic scale
 \begin{eqnarray}
\Delta^{(1)}(2 r) && = \left[ \frac{2}{(2m-2n)^{\mu \sigma}}
+ \frac{1}{(2r+1)^{\mu \sigma}} + \frac{1}{(2r-1)^{\mu \sigma}} \right]^{\frac{1}{\mu}}
\nonumber \\ &&
= \left[ 2 \Delta^{\mu}(2r) +\Delta^{\mu}(2r+1)+\Delta^{\mu}(2r-1) \right]^{\frac{1}{\mu}}
\label{deltar1}
\end{eqnarray}
in terms of the initial characteristic scale $\Delta(r) =1/r^{\sigma}$ of Eq. \ref{deltar}.

\subsection{ Iteration of the decimation procedure }

After $p$ iterations of this decimation procedure,
only spins of index $(2^p n)$ are still alive,
and the renormalized coupling between two such spins satisfy the 
renormalization rule
generalizing Eq. \ref{rgjfirst}
\begin{eqnarray}
J_{2^p n,2^p m}^{(p)} && = J_{2^p n,2^p m}^{(p-1)}
+  {\rm sgn} (J^{(p-1)}_{2^p n-2^{p-1}, 2^p n}){\rm sgn} (J^{(p-1)}_{2m-2^{p-1}, 2m})
 J_{2^p n-2^{p-1},2^p m-2^{p-1}}^{(p-1)}
\nonumber \\
&& +{\rm sgn} (J^{(p-1)}_{2^p n- 2^{p-1}, 2^p n}) J_{2^p n-2^{p-1},2^p m}^{(p-1)}
 + {\rm sgn} (J^{(p-1)}_{ 2^pm-2^{p-1}, 2^pm}) J_{2^p n,2^p m-2^{p-1}}^{(p-1)}
\label{rgjp}
\end{eqnarray}
Accordingly, these couplings remain L\'evy distributed,
and their characteristic scales as a function of the distance satisfy the recurrence
generalizing Eq. \ref{deltar1}
 \begin{eqnarray}
\left[ \Delta^{(p)}(2^p r) \right]^{\mu} &&
=  2 \left[ \Delta^{(p-1)}(2^p r) \right]^{\mu} 
+ \left[ \Delta^{(p-1)}(2^p r+2^{p-1}) \right]^{\mu} 
+ \left[ \Delta^{(p-1)}(2^p r-2^{p-1}) \right]^{\mu} 
\label{deltarp}
\end{eqnarray}
The solution reads in terms of the initial characteristic scales $\Delta(r) =1/r^{\sigma}$ of Eq. \ref{deltar} 
 \begin{eqnarray}
\left[ \Delta^{(p)}(2^p r) \right]^{\mu} &&
=  \sum_{n=-(2^p-1)}^{n=+(2^p-1)} (2^p- \vert n \vert) \Delta^{\mu}(2^p r+n)
 = \sum_{n=-(2^p-1)}^{n=+(2^p-1)} \frac{ (2^p- \vert n \vert) }{(2^p r+n)^{\mu \sigma} } 
\label{deltarpsol}
\end{eqnarray}

To see more clearly the scaling, let us now approximate 
 this discrete sum over integers $n$
by an integral over $n=2^p u$ with a continuous real variable $u \in [-1,+1]$
 \begin{eqnarray}
\left[ \Delta^{(p)}(2^p r) \right]^{\mu} &&
 \simeq \int_{-2^p}^{2^p} dn \frac{ (2^p- \vert n \vert) }{(2^p r+n)^{\mu \sigma} }
= (2^p)^{2-\mu \sigma }\int_{-1}^{1} du \frac{ (1- \vert u \vert) }{( r+u)^{\mu \sigma} }
\label{deltarpsolint}
\end{eqnarray}
In particular, on the smallest distance $R=2^p$ remaining at iteration $p$,
the characteristic scale reads
 \begin{eqnarray}
 \Delta^{(p)}( R=2^p ) 
&& \simeq R^{ \frac{2}{\mu}- \sigma } \left[ \int_{-1}^{1} du \frac{ (1- \vert u \vert) }{( 1+u)^{\mu \sigma} } \right]^{\frac{1}{\mu}} \equiv R^{\theta_{\mu}(\sigma)} C_{\mu}(\sigma)
\label{deltarpsolintfin}
\end{eqnarray}
with the droplet exponent 
 \begin{eqnarray}
\theta_{\mu}(\sigma) =\frac{2}{\mu} -\sigma
\label{thetasg}
\end{eqnarray}
that generalizes the formula of Eq. \ref{thetaLR} 
of the Gaussian case $\mu=2$ (Eq \ref{gaussian}).

Note that the decimation yields the droplet exponent of Eq. \ref{thetasg}
 only if it is positive
 \begin{eqnarray}
\theta_{\mu}(\sigma) =\frac{2}{\mu} -\sigma \geq 0
\label{thetasgpositive}
\end{eqnarray}
Indeed, if it is negative, one has to return to the discrete expression of Eq. \ref{deltarpsol}
that contains at least a term of order $O(1)$ for $n=-(2^p-1)$ even in the 
nearest-neighbor limit $\sigma=+\infty$
 \begin{eqnarray}
\left[ \Delta^{(p)}(2^p ) \right]^{\mu} &&
 = \sum_{n=-(2^p-1)}^{n=+(2^p-1)} \frac{ (2^p- \vert n \vert) }{(2^p +n)^{\mu \sigma} } \geq O(1) 
\label{deltarpsolr1}
\end{eqnarray}
So this decimation procedure is not able to reproduce negative droplet exponents,
but this problem can be overcome by introducing a strong disorder decimation,
as will be explained in the next section \ref{sec_strong}.

\subsection{Decimation with blocks of larger size $b>2$  }

\label{sec_half}

The above procedure based on blocks of size $b=2$
could be improved by the use of blocks of larger size $b>2$.
However, the droplet exponent of Eq. \ref{thetasg}
is not expected to change for the following reason.
Assume that for a system of $N$ spins, 
we use two blocks size $b=N/2$ and we find the ground states of
the internal energy in each half system.
Then the residual coupling between these two half-systems will be
\begin{eqnarray}
J_N \simeq \sum_{1 \leq i \leq \frac{N}{2}} \sum_{\frac{N}{2}+1 \leq j \leq N} J_{ij} S_i^{(1)} S_j^{(2)}
\label{j2halfs}
\end{eqnarray}
Since the spins $ S_i^{(1)} $ represent the ground state for the internal energy of the first half,
and the spins $ S_j^{(2)} $ represent the ground state for the internal energy of the second half,
they are not correlated with the couplings $J_{ij}$ between spins belonging to the two halfs,
and the Fourier transform of the residual coupling $J_N$ reads
\begin{eqnarray}
\overline{ e^{i k J_N} } && =\prod_{1 \leq i \leq \frac{N}{2}} \prod_{\frac{N}{2}+1 \leq j \leq N}   \left(\overline{  e^{i k J_{i,j} } } \right)
= \prod_{1 \leq i \leq \frac{N}{2}} \prod_{\frac{N}{2}+1 \leq j \leq N}  
  e^{ -   \frac{ \vert k \vert^{\mu}}{(j-i)^{\mu \sigma}} }
\nonumber \\
&& =   e^{ -  \vert k \vert^{\mu} \sum_{1 \leq i \leq \frac{N}{2}} \sum_{\frac{N}{2}+1 \leq j \leq N}  \frac{ 1  }{(j-i)^{\mu \sigma}} }
\label{fourierhalf}
\end{eqnarray}
so that its characteristic scale will be
\begin{eqnarray}
\Delta_N = \left[\sum_{1 \leq i \leq \frac{N}{2}} \sum_{\frac{N}{2}+1 \leq j \leq N} 
 \frac{ 1  }{(j-i)^{\mu \sigma}} \right]^{\frac{1}{\mu}} \propto N^{ \frac{2}{\mu}- \sigma }
\label{delta2halfs}
\end{eqnarray}
with the same droplet exponent as in Eq. \ref{thetasg}.
The reason why the droplet exponent does not change with the size $b$ of the box
used to make the renormalization
is also clear with this extreme case $b=N/2$ : the configurations of the spins
are determined by the couplings on the shorter scales,
while the long-ranged couplings are somewhat 'slaves' and 
are responsible for the droplet exponent. 

This extreme case $b=N/2$ also shows that the residual coupling of
Eq. \ref{j2halfs} will always contain a finite coupling $J_{i=\frac{N}{2},j=\frac{N}{2}+1}$
so that this method even for $b=N/2$
 is not able to reproduce negative droplet exponents :
the origin of this problem is that the boxes are fixed a priori independently
of the disorder realization. In the following section, 
we thus introduce a strong disorder decimation to overcome this limitation.

\section{ Strong Disorder Renormalization at zero temperature  }

\label{sec_strong}

\subsection{ Strong Disorder Decimation rule   }

As in the previous section, we wish to eliminate the odd spins $S_{2n-1}$.
But instead of choosing a priori the blocks $(S_{2n-1},S_{2n})$ as in Eqs \ref{eint}
and \ref{decimb2}, we decide to associate $S_{2n-1}$ either to its left neighbor
$S_{2n-2}$ or to its right neighbor $S_{2n}$ depending on the 
biggest coupling in absolute value between $J_{2n-2,2n-1}$ and $J_{2n-1,2n}$.
More precisely, using the Heaviside step function  
\begin{eqnarray}
\theta(x)&& = 1 \ \ {\rm if } \ \ x>0 
 \nonumber \\
\theta(x)&& = 0 \ \ {\rm if } \ \ x<0 
\label{theta}
\end{eqnarray}
the decimation rule of Eq. \ref{decimb2}
is replaced by the strong disorder decimation rule
\begin{eqnarray}
S_{2n-1 } && = 
 \theta( \vert J_{2n-1, 2n} \vert - \vert J_{2n-2, 2n-1} \vert ) {\rm sgn} (J_{2n-1, 2n}) S_{2n}
 \nonumber \\
&& +  \theta(\vert J_{2n-2, 2n-1}\vert - \vert J_{2n-1, 2n} \vert )
 {\rm sgn} (J_{2n-2, 2n-1}) S_{2n-2}
\label{rgchoice}
\end{eqnarray}
Eliminating all the odd spins with this rule, the total energy of Eq. \ref{HSGini}
becomes
\begin{eqnarray}
 E  = - \sum_n {\rm max} (\vert J_{2n-2,2n-1} \vert,\vert J_{2n-1,2n} \vert) - \sum_{-\infty \leq n <m \leq +\infty} J^{(1)}_{2n, 2m} S_{2n} S_{2n} 
\label{HSG1passtrong}
\end{eqnarray}
with the following renormalized couplings between even spins
\begin{eqnarray}
J_{2n,2 m}^{(1)} && = J_{2n,2 m} +
\nonumber \\
&&  \theta( \vert J_{2n-1, 2n} \vert -\vert J_{2n-2, 2n-1} \vert )
{\rm sgn} (J_{2n-1, 2n}) J_{2n-1,2 m}
\nonumber \\
&& + 
\theta( \vert J_{2n, 2n+1} \vert -\vert J_{2n+1, 2n+2} \vert )
{\rm sgn} (J_{2n, 2n+1}) J_{2n+1,2 m}
\nonumber \\
&&
+\theta( \vert J_{2m-1, 2m} \vert -\vert J_{2m-2, 2m-1} \vert )
 {\rm sgn} (J_{2m-1, 2m}) J_{2n,2 m-1}
\nonumber \\
&& + 
\theta( \vert J_{2m, 2m+1} \vert -\vert J_{2m+1, 2m+2} \vert )
{\rm sgn} (J_{2m, 2m+1}) J_{2n,2 m+1}
\nonumber \\
&&
+\theta( \vert J_{2n-1, 2n} \vert -\vert J_{2n-2, 2n-1} \vert )
\theta( \vert J_{2m-1, 2m} \vert -\vert J_{2m-2, 2m-1} \vert )
 {\rm sgn} (J_{2n-1, 2n} J_{2m-1, 2m}) J_{2n-1,2 m-1}
\nonumber \\
&&
+ \theta( \vert J_{2n-1, 2n} \vert -\vert J_{2n-2, 2n-1} \vert )
\theta( \vert J_{2m, 2m+1} \vert -\vert J_{2m+1, 2m+2} \vert )
 {\rm sgn} (J_{2n-1, 2n} J_{2m, 2m+1}) J_{2n-1,2 m+1}
\nonumber \\
&&
+\theta( \vert J_{2n, 2n+1} \vert -\vert J_{2n+1, 2n+2} \vert )
\theta( \vert J_{2m-1, 2m} \vert -\vert J_{2m-2, 2m-1} \vert )
{\rm sgn} (J_{2n, 2n+1} J_{2m-1, 2m} )J_{2n+1,2 m-1}
\nonumber \\
&&
+\theta( \vert J_{2n, 2n+1} \vert -\vert J_{2n+1, 2n+2} \vert )
\theta( \vert J_{2m, 2m+1} \vert -\vert J_{2m+1, 2m+2} \vert )
{\rm sgn} (J_{2n, 2n+1} J_{2m, 2m+1} )J_{2n+1,2 m+1}
\label{rgjfirstchoice}
\end{eqnarray}
This renormalization rule is of course much more complicated that Eq \ref{rgjfirst},
but the physical meaning is clear. The rule of Eq. \ref{rgchoice} means
that the correlated cluster which is constructed around
the even spin $S_{2n}$ does not have the fixed size of $b=2$ spins (as in the rule
of Eq. \ref{decimb2} studied in the previous section)
but can have for size $b_{2n}=1$ (if its two neighbors $S_{2n-1}$ and $S_{2n+1}$
are linked to their other neighbors via the rule of Eq. \ref{rgchoice}),
or $b_{2n}=2$ (if only one of its two neighbors $S_{2n-1}$ and $S_{2n+1}$ 
is linked to $S_{2n}$ via the rule of Eq. \ref{rgchoice})
or $b_{2n}=3$ (if its two neighbors $S_{2n-1}$ and $S_{2n+1}$
are linked to $S_{2n}$ via the rule of Eq. \ref{rgchoice}).
As a consequence, the renormalized coupling $J_{2n,2 m}^{(1)}$ of Eq. \ref{rgjfirstchoice}
between two such
clusters is a sum over a fluctuating number $n_s=b_{2n}b_{2m}$ of couplings
with the possible values $ns=1,2,3,4,6,9$ for $n<m+1$.

\subsection{ Exactness for the nearest-neighbor spin-glass chain ($\sigma=+\infty$)  }

\label{sec_nn1d}

The nearest-neighbor spin-glass chain
\begin{eqnarray}
 E  = - \sum_{n} J_{n ,n+1} S_n S_{n+1}
\label{HSGininn}
\end{eqnarray}
corresponds to the long-ranged model of Eqs \ref{HSGini} and
\ref{defjijini} in the limit $\sigma=+\infty$.
In this limit, the renormalization rule of Eq. \ref{rgjfirstchoice} reduces to
\begin{eqnarray}
J_{2n,2n+2}^{(1)} && = 
\theta( \vert J_{2n, 2n+1} \vert -\vert J_{2n+1, 2n+2} \vert )
{\rm sgn} (J_{2n, 2n+1}) J_{2n+1,2n+2}
\nonumber \\
&&
+\theta( \vert J_{2n+1, 2n+2} \vert -\vert J_{2n, 2n+1} \vert )
 {\rm sgn} (J_{2n+1, 2n+2}) J_{2n,2n+1}
\nonumber \\
&& =  {\rm sgn} (J_{2n, 2n+1} J_{2n+1,2n+2})
 {\rm min} (\vert J_{2n,2n+1} \vert;\vert J_{2n+1,2n+2} \vert)
\label{nearestneighborfirstchoice}
\end{eqnarray}
More generally, after $p$ iterations, the renormalized couplings existing between
two sites separated by the distance $2^p$ reads
\begin{eqnarray}
J_{2^pn,2^pn+2^p}^{(1)} 
&& =  {\rm sgn} ( \prod_{j=1}^{2^p} J_{2n+j-1, 2n+j} )
 {\rm min}_{1 \leq j \leq 2^p} (\vert J_{2n+j-1,2n+j} \vert \vert)
\label{nearestneighborfirstchoicep}
\end{eqnarray}
in agreement with the exact result \cite{bray_moore} :
on scale $L=2^p$, the absolute value of the renormalized coupling 
is determined by the minimal value in absolute value of the $L=2^p$
initial couplings existing between them \cite{bray_moore}
\begin{eqnarray}
\vert J_{2^pn,2^pn+2^p}^{(1)} \vert
&& =   {\rm min}_{1 \leq j \leq 2^p} (\vert J_{2n+j-1,2n+j} \vert \vert) \propto \frac{1}{2^p} = \frac{1}{L}
\label{nearestneighborfirstchoicepabs}
\end{eqnarray}
corresponding to the exact droplet exponent of Eq. \ref{thetaSRd1}
 \begin{eqnarray}
\theta^{SR}_{\mu}(\sigma=+\infty) =-1
\label{thetannmu}
\end{eqnarray}
Note that this result is valid both for Gaussian ($\mu=2$) and for L\'evy couplings
($1<\mu<2$),
since the only important property of the initial distribution leading to the scaling
of Eq. \ref{nearestneighborfirstchoicep} is the finite weight near zero coupling $P(J=0)$.
So this nearest-neighbor case, even if it is trivial, shows that the strong disorder rule
of Eq. \ref{rgchoice} is exact in the limit $\sigma=+\infty$, and that it is able
to yield negative droplet exponent, in contrast to the block decimation of the previous section.

\subsection{ Exactness for the Migdal-Kadanoff approximation  }

The Migdal-Kadanoff approximation 
amounts to replace the hypercubic lattice by
a hierarchical fractal diamond lattice \cite{MKRG,berker,hierarchical}
 which is constructed recursively as follows.
Two boundary spins $S_A$ and $S_B$ are linked by $K$ branches
each branch containing two links and a middle spin $S_i$, so that the corresponding energy reads
 \begin{eqnarray}
E_{S_A,S_B}= -\sum_{i=1}^K \left[ J_{A,i} S_A S_i +J_{B,i} S_B S_i  \right]
\label{ediamond}
\end{eqnarray}
The strong disorder rule of Eq. \ref{rgchoice} reads for each interior spin $S_i$
\begin{eqnarray}
S_{i} && = 
 \theta( \vert J_{A,i} \vert - \vert J_{B,i} \vert ) {\rm sgn} (J_{A,i}) S_{A}
 \nonumber \\
&& +   \theta( \vert J_{B,i} \vert - \vert J_{A,i} \vert ) {\rm sgn} (J_{B,i}) S_{B}
\label{rgchoicediamond}
\end{eqnarray}
Plugging this expression into Eq. \ref{ediamond} yields
 \begin{eqnarray}
E_{S_A,S_B} && = -\sum_{i=1}^K {\rm max} \left[ 
\vert J_{A,i} \vert , \vert J_{B,i} \vert \right]
 -  J_{AB}^R S_A S_B
\label{ediamondrg}
\end{eqnarray}
with the renormalized coupling
 \begin{eqnarray}
J_{AB}^R = \sum_{i=1}^K
{\rm sign} ( J_{A,i} J_{B,i})  {\rm min} \left[ \vert J_{A,i} \vert ;\vert J_{B,i} \vert \right]
\label{jdiamondrg}
\end{eqnarray}
which coincides with the zero-temperature limit of the 
Migdal-Kadanoff RG rules for spin-glasses, obtained by exact recursions on partition functions
 \cite{southern_young,young_St,BM_MK,mckay,Gardnersg,banavar,bokil,muriel,thill,boettcher_mk,us_tail,jorg,us_sgferro}.

\subsection{ Numerical results for the Gaussian Long-ranged spin-glass  }

For the Gaussian case $\mu=2$, we have applied numerically the strong disorder 
renormalization rule of Eq. \ref{rgjfirstchoice} for chains containing 
initially $N=2^{13}=8192$ spins (the number of initial couplings being $N(N-1)/2 \simeq 33.10^6$),
with periodic boundary conditions, 
for the following values of the long-ranged power $\sigma=0.55 ; 0.75 ; 0.87 ; 1.25 ; 1.5 ; 2 ; 3 ; 10 $.
Our numerical results for the variance of the renormalized couplings $J^{(p)}$
at iterations $1 \leq p \leq 13$ on the corresponding elementary length scale $2^p$
are in agreement with the theoretical expectation of Eq. \ref{thetaLRd1},
even in the region $\sigma >1$ where the droplet exponent is negative.

Our conclusion is thus that the strong disorder decimation yields 
the same exponent as the box decimation (Eq. \ref{thetasg})
in the region $1/2<\sigma<1$ where the droplet exponent is positive,
but is also able to reproduce the correct negative droplet exponent in the region $\sigma>1$.

\subsection{ Difference with strong disorder renormalization for quantum spin models }

As a final remark, we should stress the difference with the 
strong disorder renormalization method (see \cite{review_strong} for a review) that has been developed for
disordered {\it  quantum } short-ranged spin models either in $d=1$ \cite{fisher} or in $d=2,3,4$ 
 \cite{motrunich,fisherreview,lin,karevski,lin07,yu,kovacsstrip,kovacs2d,kovacs3d,kovacsentropy,kovacsreview}.
In these quantum spin models, the idea is to decimate the strongest coupling $J_{max}$ remaining in the whole system :  the renormalized couplings obtained via second order perturbation theory of quantum mechanics
are typically much weaker than the decimated coupling $J_{max}$, so that the procedure is consistent and the typical renormalized couplings {\it decays }  as $J_L^{typ} \propto e^{- L^{\psi}}$. This is thus completely different from the problem of {\it classical } spin-glasses considered in the present paper, where the interesting spin-glass phases are governed by a positive droplet exponent
$\theta$, so that the typical coupling {\it grows} with the scale $J_L \propto L^{\theta}$. This is why here we have included the strong disorder rule of Eq. \ref{rgjfirstchoice}
within the more traditional decimation framework that fixes the length scale.

\section{ Stability with respect to an external magnetic field}

\label{sec_magnetic}

The stability of the spin-glass phase with respect to an external magnetic field,
i.e. the presence of an Almeida-Thouless line in the phase diagram,
is one of the most controversial issue in spin-glass theory, 
and there are, at least, three opinions for short-ranged spin-glasses :

(i) in the Fisher-Huse droplet scaling theory \cite{fisher_huse},
there is no Almeida-Thouless line in any finite dimension $d$.

(ii) in the Bray-Moore replica analysis \cite{MB_AT},
the Almeida-Thouless line exists only above the upper critical dimension $d \geq d_u=6$.

(iii)  in the Parisi-Temesvari replica analysis \cite{PT_AT},
the Almeida-Thouless line exists also below the upper critical dimension $d_u=6$.

From the point of view of the droplet scaling theory \cite{fisher_huse}
that we follow in the present paper,
the stability of the spin-glass with respect to a small magnetic field $H$
can be predicted via the following Imry-Ma scaling argument : 
on a linear scale $L$, the renormalized sin-glass coupling $J_L$ 
of Eq. \ref{deftheta}
has to be compared with the perturbation of order
\begin{eqnarray}
 \delta_H(L) \propto H L^{1/2}
\label{Hper}
\end{eqnarray}
induced by the field $H$ coupled to the random magnetization of order $ L^{1/2}$.
As a consequence, the spin-glass will be unstable with respect to the magnetic field
if $\theta<1/2$, and stable if $\theta>1/2$.
For the Gaussian case $\mu=2$, 
this Imry-Ma argument predicts that the Gaussian long-ranged 
spin-glass is always unstable in the extensive region
$\mu >1/2$, whereas the L\'evy long-ranged spin-glass is stable in the region
\begin{eqnarray}
 \frac{1}{\mu} < \sigma <  \frac{2}{\mu}-\frac{1}{2}
\label{stabB}
\end{eqnarray}

\section{ Statistics of the ground state energy}

\label{sec_energy}

\subsection{ Reminder }

The statistics over samples of the ground state energy in spin-glasses 
has been much studied recently (see
\cite{andreanov,Bou_Krz_Mar,pala_gs,aspelmeier_MY,Katz_gs,Katz_guiding,aspelmeier_BMM,boettcher_gs,us_tails,us_matching} and references therein) with the following conclusions :

(i) the averaged value over samples of the ground state energy reads
\begin{eqnarray}
\overline{ E^{GS}(N=L^d) }  \simeq L^{d} e_0+ L^{\theta_{shift}} e_1+...
\label{e0av}
\end{eqnarray}
The first term $L^{d} e_0$ is the extensive contribution in the number $N=L^d$ number of spins,
whereas the second term $L^{\theta_{shift}} e_1$ represents the leading correction to
extensivity.

(ii) the fluctuations around this averaged value are governed by some fluctuation exponent
$g$
\begin{eqnarray}
E^{GS}(N)- \overline{ E^{GS}(N) }  \simeq N^{g} u +...
\label{e0fluct}
\end{eqnarray}
where $u$ is an $O(1)$ random variable of zero mean $\overline{u}=0$ 
distributed with some distribution $G(u)$.

For Gaussian spin-glasses in finite dimension $d$, it has been proven \cite{aizenman_wehr}
that 
\begin{eqnarray}
g^{SR}=\frac{1}{2}
\label{gSR}
\end{eqnarray}
 that the distribution $G(u)$ of the rescaled variable $u$ is simply Gaussian.

But the shift-exponent of Eq. \ref{e0av} remains nevertheless non-trivial
 and is expected to coincide with the droplet exponent $¿theta$ \cite{Bou_Krz_Mar}
\begin{eqnarray}
\theta_{shift}= \theta
\label{thetashift}
\end{eqnarray}

For the long-ranged spin-glass with Gaussian distribution, the statistics 
of the ground state energy has been studied numerically  \cite{KY,KKLH,KKLJH}
with various conclusions : 
the numerical results for the width exponent $g$ obtained in Reference \cite{KY} 
are clearly below $g=1/2$ in the region $\sigma \to (1/2)^+$,
whereas the authors of References \cite{KKLH,KKLJH} have concluded that
the asymptotic distribution $G(u)$ is Gaussian in the whole extensive region $\sigma>1/2$.
It is thus interesting to analyze the result given by the renormalization procedures
described in the previous sections.

\subsection{ Analysis within the block decimation }

Within the box decimation of size $b=2$ described in section \ref{sec_box},
the obtained ground state energy for a system of length $N=2^{p_{max}}$
can be decomposed into the sum of
the energies $E_N^{(p)}$ gained at each RG iteration $p$ by the satisfaction of 
half of the renormalized bonds corresponding to the elementary length scale $2^{p-1}$
of this iteration
\begin{eqnarray}
 E_N^{GS}  =  \sum_{p=1}^{p_{max}} E_N^{(p)}
\label{egsep}
\end{eqnarray}
We have already seen in Eq. \ref{HSG1pas}
that the energy $E_N^{(p=1)}$ gained at the first decimation $p=1$ reads
\begin{eqnarray}
 E_N^{(p=1)}  = - \sum_{1 \leq n \leq \frac{N}{2}} \vert J_{2n-1,2n} \vert  
\label{HSG1pase1}
\end{eqnarray}
so that we may rewrite it as the sum over $(N/2)$ independent coupling 
 of the initial distribution on the elementary length scale $2^0=1$
\begin{eqnarray}
E_N^{(p=1)}   = -  \sum_{i_0=1}^{\frac{N}{2}} \vert J^{(0)}_{i_0} \vert
\label{e1}
\end{eqnarray}
More generally, at the iteration $p$, the energy $E_N^{(p)}$ gained by the satisfaction of 
half of the renormalized bond of generation $(p-1)$ 
 on the elementary scale $2^{p-1}$ reads
\begin{eqnarray}
E_N^{(p)}   = -  \sum_{i_p=1}^{\frac{N}{2^p}} \vert J^{(p-1)}_{i_p} \vert
\label{ep}
\end{eqnarray}
up to the last renormalization scale $p_{max}= \frac{\ln N}{\ln 2}$, where 
the energy $E_N^{(p_{max})}$ is gained by the satisfaction of 
the single renormalized bond of generation $(p_{max}-1)$
 on the elementary scale $2^{p_{max}-1}=N/2$
coupling the two halfs of the system
\begin{eqnarray}
E_N^{(p_{max})}   =  - \vert J^{(p_{max}-1)}_{1} \vert
\label{epmax}
\end{eqnarray}

\subsubsection{ Analysis of the averaged value }

In the region $1<\mu \leq 2$ that we consider (Eq \ref{regionmusigma}),
 the L\'evy distribution of index $\mu$
(Eqs \ref{levy} and \ref{levytail})
has a finite first moment in absolute value
 \begin{eqnarray}
B_{\mu} \equiv \int_{-\infty}^{+\infty} d \epsilon \vert \epsilon \vert L_{\mu}(\epsilon) < +\infty
\label{levymom1}
\end{eqnarray}
As a consequence, at scale $p$, using Eq. \ref{deltarpsolintfin},
one obtains that the average value of the absolute coupling reads
\begin{eqnarray} 
\overline {\vert J^{(p)} \vert} = B_{\mu} \Delta^{(p)}(2^p) \simeq  B_{\mu} C_{\mu}(\sigma)
 (2^p)^{\theta_{\mu}(\sigma)}
\label{jpabsav}
\end{eqnarray}
So the contribution of the $p$ iteration reads (Eq \ref{ep})
\begin{eqnarray}
 \overline{ E_N^{(p)} }  = -  \frac{N}{2^p} \overline{ \vert J^{(p-1)} \vert }
\simeq - \frac{N}{2^p}  B_{\mu} C_{\mu}(\sigma) (2^p)^{\theta_{\mu}(\sigma)}
\label{epav}
\end{eqnarray}
Finally, the averaged value of the groundstate energy obtained by
the box decimation reads (Eq \ref{egsep})
\begin{eqnarray}
 \overline{E_N^{GS}}  =  - \sum_{p=1}^{p_{max}} \overline{ E_N^{(p)} }
= - N B_{\mu} C_{\mu}(\sigma)  \sum_{p=1}^{p_{max}= \frac{\ln N}{\ln 2}} (2^p)^{\theta_{\mu}(\sigma)-1}
\label{egsepav}
\end{eqnarray}
In the region that we consider (Eq. \ref{regionmusigma}),
one has $\theta_{\mu}(\sigma)<1 $, so that the
the last geometric sum  is always convergent and the averaged ground state energy is extensive
in the number $N$ of spins as expected.
In addition, one obtains that the contribution of the last decimation $p_{max}=\frac{\ln N}{\ln 2}$ of Eq. (\ref{epmax} and \ref{epav}) reads
\begin{eqnarray}
\overline{E_N^{(p_{max})} }  =  - \overline{ \vert J^{(p_{max}-1)} \vert }
= - B_{\mu} C_{\mu}(\sigma) N^{\theta_{\mu}(\sigma)}
\label{epmaxav}
\end{eqnarray}
so that the exponent $\theta_{shift}$ governing the leading correction to extensivity
(Eq \ref{e0av})
indeed coincides with the droplet exponent $ \theta_{\mu}(\sigma)$ as expected 
in general (Eq. \ref{thetashift}).

\subsubsection{ Distribution around the averaged value in the Gaussian case $\mu=2$ }

In the Gaussian case $\mu=2$ (Eq. \ref{gaussian}), 
the distribution of the coupling $J^{(p)}$ at iteration $p$ on scale $2^p$
is also Gaussian 
 \begin{eqnarray}
P_p(J^{(p)}) =   \frac{1}{\Delta^{(p)}(2^p) \sqrt{4 \pi} }
 e^{- \frac{1}{4} \left(\frac{(J^{(p)})}{\Delta^{(p)}(2^p)}\right)^2}
\label{gaussianp}
\end{eqnarray}
of characteristic scale (Eq. \ref{deltarpsolintfin})
 \begin{eqnarray}
\Delta^{(p)}(2^p) = (2^p)^{\theta_{2}(\sigma)} C_2(\sigma)
\label{gaussianpdelta}
\end{eqnarray}
so that the variance of the absolute value of the coupling reads
\begin{eqnarray}
 \overline{ (J^{(p-1)} )^2 } - ( \overline{\vert J^{(p-1)} \vert  } )^2 = \left(2-\frac{4}{\pi} \right) \Delta^{(p)}(2^p)  = \left(2-\frac{4}{\pi} \right)C_2(\sigma) (2^p)^{\theta_{2}(\sigma)}
\label{varjp}
\end{eqnarray}
For finite $p$, the Central Limit Theorem yields
that the energy $E_N^{(p)}$ gained at the decimation $p$ 
will be Gaussian distributed, with an averaged given by Eq. \ref{epav}),
and with a variance given by
\begin{eqnarray}
\overline{ (E_N^{(p)})^2 } - ( \overline{ E_N^{(p)} } )^2   =  \frac{N}{2^p}
\left[ \overline{ (J^{(p-1)} )^2 } - ( \overline{\vert J^{(p-1)} \vert  } )^2  \right]
= \frac{N}{2^p} \left(2-\frac{4}{\pi} \right)C_2(\sigma) (2^p)^{\theta_{2}(\sigma)}
\label{epvar}
\end{eqnarray}
As a consequence,  the Central Limit Theorem yields
that the distribution of the ground state energy (Eq. \ref{egsep})
will also be Gaussian,  with an averaged given by Eq. \ref{egsepav}
and with a variance given by
\begin{eqnarray}
\overline{ (E_N^{GS})^2 } - ( \overline{ E_N^{GS} } )^2   =
  \sum_{p=1}^{p_{max}} \left[ \overline{ (E_N^{(p)})^2 } - ( \overline{ E_N^{(p)} } )^2  \right]
= N \left(2-\frac{4}{\pi} \right)C_2(\sigma) \sum_{p=1}^{p_{max}= \frac{\ln N}{\ln 2}}
 (2^p)^{\theta_{2}(\sigma)-1}
\label{etotvar}
\end{eqnarray}
so that the fluctuation exponent $g$ of Eq. \ref{e0fluct} is simply
\begin{eqnarray}
g_{\mu=2}=\frac{1}{2}
\label{ggauss}
\end{eqnarray}
 in agreement with the result for short-ranged Gaussian spin-glasses in any finite dimension $d$
 (Eq. \ref{gSR}).

\subsubsection{Distribution around the averaged value in the L\'evy case $1<\mu<2$  }

 In the L\'evy case $1<\mu<2$, the distribution of the 
coupling $J^{(p)}$ at iteration $p$ on scale $2^p$ displays the power-law tail
of Eq. \ref{levytail}
 \begin{eqnarray}
P_p(J^{(p)})  \opsimeq_{ J^{(p)}  \to \pm \infty} \frac{A_{\mu} \left[\Delta^{(p)}(2^p) \right]^{\mu} }{\vert J^{(r)} \vert^{1+\mu}}
\label{levytailr}
\end{eqnarray}
of characteristic scale (Eq. \ref{deltarpsolintfin})
 \begin{eqnarray}
\Delta^{(p)}(2^p) = (2^p)^{\theta_{\mu}(\sigma)} C_{\mu}(\sigma)
\label{levypdelta}
\end{eqnarray}
As a consequence, from the theory of the addition of random L\'evy variables,
the energy $E_N^{(p)}$ gained at the decimation $p$ (Eq. \ref{ep})
will display the following power-law negative tail
 \begin{eqnarray}
Q_p(E_N^{(p)})  \opsimeq_{ E_N^{(p)}  \to - \infty} \frac{A_{\mu} \frac{N}{2^p} \left[\Delta^{(p)}(2^p) \right]^{\mu} }
{ (-E_N^{(p)} ) ^{1+\mu}}
\label{levytailer}
\end{eqnarray}
and the ground state energy of Eq. \ref{egsep}
the following power-law negative tail
 \begin{eqnarray}
{\cal P}(E_N^{GS})  \opsimeq_{ E_N^{GS}  \to - \infty} \frac{A_{\mu} \left[\Delta_N^{GS} \right]^{\mu} }
{ (-E_N^{GS} ) ^{1+\mu}}
\label{levytaileegs}
\end{eqnarray}
of characteristic scale $\Delta^{GS}_N $ such that 
 \begin{eqnarray}
\left[ \Delta_N^{GS} \right]^{\mu} && =  \sum_{p=1}^{p_{max}} \frac{N}{2^p} \left[\Delta^{(p)}(2^p) \right]^{\mu} 
= \sum_{p=1}^{p_{max}} \frac{N}{2^p} \left[  (2^p)^{\theta_{\mu}(\sigma)} C_{\mu}(\sigma) \right]^{\mu}
\nonumber \\
&& = N C_{\mu}^{\mu}(\sigma) \sum_{p=1}^{p_{max}=\frac{\ln N}{\ln 2}}   (2^p)^{\mu\theta_{\mu}(\sigma)-1} 
\label{deltaGS}
\end{eqnarray}
From the expression of the droplet exponent of Eq. \ref{thetasg},
it is clear that the exponent $\mu\theta_{\mu}(\sigma)-1=1-\mu \sigma$
is always negative in the region that we consider (Eq. \ref{regionmusigma}),
so that the final geometric sum of Eq. \ref{deltaGS} is always convergent.
As a consequence, the characteristic scale $\Delta^{GS}$ of the ground-state energy
in its negative tail is of order 
 \begin{eqnarray}
 \Delta^{GS}_N \propto N^{\frac{1}{\mu}}
\label{deltagsmu}
\end{eqnarray}
so that the fluctuation exponent 
 $g$ of Eq. \ref{e0fluct} is simply
\begin{eqnarray}
g_{\mu}=\frac{1}{\mu}
\label{glevy}
\end{eqnarray}
that generalizes the value of Eq. \ref{ggauss} concerning the Gaussian case $\mu=2$.

\subsection{ Analysis for the strong disorder decimation } 

For the strong disorder decimation, the energy $E_N^{(1)}$ gained by the first iteration $p=1$
reads (Eq \ref{HSG1passtrong})
\begin{eqnarray}
 E_N^{(1)}  = - \sum_{1 \leq n \leq \frac{N}{2}} {\rm max} (\vert J_{2n-2,2n-1} \vert,\vert J_{2n-1,2n} \vert) 
\label{e1strong}
\end{eqnarray}
So instead of the sum over $(N/2)$ independent coupling for the box decimation (Eq \ref{e1}),
it is given by the sum over $(N/2)$ independent variables,
each variable being the minimum of two independent couplings
\begin{eqnarray}
E_N^{(p=1)}   = -  \sum_{i_0=1}^{\frac{N}{2}} {\rm max} (\vert J^{(0)}_{i_0 (a)} \vert,\vert J^{(0)}_{i_0 (b)})
\label{e1strongbis}
\end{eqnarray}
As a consequence, we do not expect any changes in the scalings discussed above, but only in the constants.
For instance, the intensive energy $e_0$ of Eq. \ref{e0av} will be lower for the
strong disorder decimation than with the box decimation.

\section{ Renormalization for the Dyson hierarchical spin-glass model  } 

\label{sec_dyson}

In the field of long ranged models, it is also interesting to consider
their Dyson hierarchical analogs, where real space renormalization are easier to define as a consequence of the hierarchical structure.
After many works concerning the Dyson hierarchical ferromagnetic Ising model
 \cite{dyson} by both mathematicians
\cite{bleher,gallavotti,book,jona} and physicists \cite{baker,mcguire,Kim,Kim77,us_dysonferrodyn},
various Dyson hierarchical disordered systems 
 have been recently studied,
in particular Anderson localization models \cite{bovier,molchanov,krit,kuttruf,fyodorov,EBetOG,fyodorovbis,us_dysonloc}, random fields Ising models
\cite{randomfield,us_aval} and spin-glasses \cite{castel_etal,castel_parisi,castel,angelini}. 
In the following, we thus apply the zero-temperature renormalization
to the Dyson hierarchical spin-glass in order to stress the similarities
and differences with respect to the long-ranged model discussed in previous sections.

\subsection{ Definition of the Dyson hierarchical spin-glass model  } 

The Dyson hierarchical spin-glass model of $2^N$ spins
is defined recursively as follows \cite{castel_etal,castel_parisi,castel,angelini}
\begin{eqnarray}
H_{N}(S_1,S_2,...,S_{2^N}) &&  = H_{N-1}^{(a)}(S_1,S_2,...,S_{2^{N-1}})
+H_{N-1}^{(b)}(S_{2^{N-1}+1},S_{2^{N-1}+2},...,S_{2^{N}})
\nonumber \\
&&  -  \sum_{i=1}^{2^{N-1}} \sum_{j=2^{N-1}+1}^{2^N} J_{N-1}(i,j) S_i S_j 
\label{sgDyson}
\end{eqnarray}
(where the notation $H_{N-1}^{(a)} $ and $H_{N-1}^{(b)} $ means that
these two Hamiltonians are two independent realizations for the two half-systems
before the introduction of the couplings of the second line).
The first terms for $N=1$ and $N=2$ reads
\begin{eqnarray}
H_{1}(S_1,S_2)  && = -J_0(1,2) S_1 S_2
\nonumber \\
H_{2}(S_1,S_2,S_3,S_4)  && = -J_0(1,2) S_1 S_2-J_0(3,4) S_3 S_4 \nonumber \\
&&
- J_1(1,3) S_1 S_3 - J_1(1,4) S_1 S_4
- J_1(2,3) S_2 S_3 - J_1(2,4) S_2 S_4
\label{sgDyson1}
\end{eqnarray}
At generation $n$, associated to the length scale $L_n=2^n$,
the couplings $J_n(i,j) $ are 
independent L\'evy variables of zero-mean and characteristic scale $\Delta_n$,
with the following exponential decay with the number $n$ of generations,
in order to mimic the power-law decay of Eq. \ref{deltar}
 with respect to the length scale $L_n=2^{n}$
\begin{eqnarray}
\Delta_n = 2^{-n \sigma} =  \frac{1}{L_n^{\sigma}} 
\label{jndysonsg}
\end{eqnarray}
As a consequence of this hierarchical structure, the block decimation
of section \ref{sec_box} is directly appropriate for the Dyson model, 
since the blocks are already built in the definition of the model,
and one does not need to introduce the strong disorder version of section
 \ref{sec_strong}. 

\subsection{ Decimation using blocks of size $b=2$  } 

The decimation of all odd spins using the rule of Eq. \ref{decimb2}
amounts to satisfy all the couplings of generation zero $J_0(i,j)$.
Then the renormalized couplings between the remaining even spins
associated to generation $n \geq 1$ read
\begin{eqnarray}
J_n^{(1)}(2i,2j) && = J_{n}(2i,2j)
 + {\rm sgn} [J_0(2i-1, 2i){\rm sgn} (J_{0}(2j-1, 2j)] J_n(2i-1,2j-1)
\nonumber \\
&&  +{\rm sgn} [J_0(2i-1, 2i)] J_n(2i-1,2j)
+ {\rm sgn} [J_0(2j-1, 2j)] J_n(2i,2j-1)
\label{rgjfirstdyson}
\end{eqnarray}
that replaces Eq. \ref{rgjfirst} concerning the non-hierarchical case.
As a consequence, the RG rule for the Fourier transform reads
(instead of Eq. \ref{fourierrgjfirst})
\begin{eqnarray}
\overline{ e^{i k J_n^{(1)}(2i,2j) }} && =\left(  \overline{  e^{i k J_n(2i,2j)  } } \right)
\left( \overline{ e^{i k J_n(2i-1,2j-1) } } \right)
\left(\overline{ e^{i k  J_n(2i-1,2j) }} \right)
\left( \overline{ e^{i k  J_n(2i,2j-1) }} \right)
\nonumber \\ && = e^{ - 4 \vert k \vert^{\mu} \Delta_n^{\mu} }
\label{fourierrgjfirstdyson}
\end{eqnarray}
so that the renormalized characteristic scale of the L\'evy distribution 
evolves simply as
(instead of Eq. \ref{deltar1})
 \begin{eqnarray}
\Delta^{(1)}_n && = \left[  4 \Delta_n^{\mu} \right]^{\frac{1}{\mu}} = 4^{\frac{1}{\mu}}
\Delta_n
\label{deltar1dyson}
\end{eqnarray}
in terms of the initial characteristic scale $\Delta_n $ of Eq. \ref{jndysonsg}.

This procedure can be then straightforwardly iterated :
after $p$ iterations where only spins of index $(2^p i)$ are still alive,
the characteristic scale $\Delta^{(p)}_n $ of the couplings of generation $n \geq p$
after $p$ renormalization steps reads (instead of Eq. \ref{deltarpsol})
 \begin{eqnarray}
\Delta^{(p)}_n && = \left( 4^{\frac{1}{\mu}} \right)^p \Delta_n
\label{deltarpdyson}
\end{eqnarray}
In particular for the smallest remaining generation $n=p$ associated to the
length scale $2^p$, one obtains using the initial value of Eq. \ref{jndysonsg}
 \begin{eqnarray}
\Delta^{(p)}_p && = 2^{\frac{2p}{\mu}}  \Delta_p
 = 2^{ p \left(\frac{2}{\mu}-\sigma \right)} = (2^p)^{\theta_{\mu}(\sigma)}
\label{deltarppdyson}
\end{eqnarray}
with the same droplet exponent $\theta_{\mu}(\sigma)=\frac{2}{\mu}-\sigma$
as in Eq. \ref{thetasg}, but the domain of validity is different :
here the formula $\theta_{\mu}(\sigma)=\frac{2}{\mu}-\sigma $ is always valid,
 even when the droplet exponent is negative $\theta_{\mu}(\sigma)=\frac{2}{\mu}-\sigma \leq 0 $ (in contrast to Eqs  \ref{thetasgpositive}
and \ref{deltarpsolr1} concerning the non-hierarchical case).

\subsection{ Decimation using blocks of size $b>2$  } 

To see whether results change if blocks of larger size $b>2$ are used
for the decimation,
let us consider the extremal case of two blocks of size $b=2^{N-1}$.
The residual coupling between the ground states of the two halfs
reads (instead of Eq. \ref{j2halfs})
\begin{eqnarray}
J_{N-1}^{(R)} = \sum_{i=1}^{2^{N-1}} \sum_{j=2^{N-1}+1}^{2^N} J_{N-1}(i,j)
 S_i^{(a)} S_j^{(b)}
\label{j2halfsdyson}
\end{eqnarray}
where $S_i^{(a)} $ and $S_j^{(b)}$ are the ground states of the two independent Hamiltonians $H_{N-1}^{(a)}$ and $H_{N-1}^{(b)}$ of Eq. \ref{sgDyson}.
As a consequence, its characteristic scale
 will be (instead of Eq. \ref{delta2halfs})
\begin{eqnarray}
\Delta_{N-1}^R = \left[ 
\sum_{i=1}^{2^{N-1}} \sum_{j=2^{N-1}+1}^{2^N} \Delta_{N-1}^{\mu} \right]^{\frac{1}{\mu}}
= 2^{\frac{ 2(N-1) }{\mu} } \Delta_{N-1}
= (2^{N-1})^{ \frac{2}{\mu}- \sigma } = L_{N-1}^{\theta_{\mu}(\sigma)}
\label{delta2halfsdyson}
\end{eqnarray}
with the same droplet exponent  $\theta_{\mu}(\sigma)=\frac{2}{\mu}-\sigma$
as in Eq. \ref{deltarppdyson} obtained with blocks of size $b=2$.
This analysis also shows that the formula remains even valid in the domain 
$\theta_{\mu}(\sigma)=\frac{2}{\mu}-\sigma \leq -1= \theta^{SR}(d=1) $
(in contrast to Eqs \ref{thetaLR} and \ref{thetaLRd1}), because here
there is no competition with the short-ranged droplet exponent
(the short-ranged model is {\it not} included in the Dyson
hierarchical model, whereas it is included in the non-hierarchical model).

In summary, the Dyson hierarchical spin-glass model 
leads to very simple zero-temperature decimation rules 
and to the droplet exponent $\theta_{\mu}(\sigma)=\frac{2}{\mu}-\sigma $
even in the region where it is negative.

\section{ Conclusion}

\label{sec_conclusion}

In this paper, we have considered the one-dimensional long-ranged Ising spin-glass with random couplings decaying with the distance $r$ as $J(r) \sim r^{-\sigma}$ and distributed with the L\'evy symmetric stable distribution of index $1 <\mu \leq 2$ (including the usual Gaussian case $\mu=2$), in the region $\sigma>1/\mu$ where the energy is extensive.  We have analyzed two real space renormalization procedures at zero temperature. The simple box decimation leads to explicit calculations, but gives the correct droplet exponent only if it is positive. The strong disorder decimation is better since it is also able to reproduce negative droplet exponents,
and it becomes exact in the Migdal-Kadanoff approximation.

The known formula of Eq. \ref{thetaLRd1} for the Gaussian case $\mu=2$ reads for the L\'evy case
of index $\mu$
\begin{eqnarray}
\theta_{\mu}(\sigma)=\frac{2}{\mu}-\sigma
\end{eqnarray}
with the consequence that there exists a non-empty region (Eq \ref{stabB}) where the spin-glass phase is expected to be stable with respect to a small magnetic field.

We have also analyzed in detail the consequences for the
statistics of the ground state energy $E_L^{GS}$ over disordered samples.
We have obtained that the droplet exponent $\theta_{\mu}(\sigma) $ governs the leading correction to extensivity of the averaged value $\overline{E_L^{GS}} \simeq L e_0 +L^{\theta_{\mu}(\sigma)} e_1$, as a consequence of the last RG step at the biggest scale, whereas the characteristic scale of the fluctuations around this average is of order $L^{\frac{1}{\mu}}$ as a consequence of the extensive number of couplings on the short scales. The rescaled variable $u=(E_L^{GS}-\overline{E_L^{GS}})/L^{\frac{1}{\mu}}$ is then Gaussian for initial couplings with a finite variance, or displays the negative power-law tail in $1/(-u)^{1+\mu}$ for $u \to -\infty$ in the L\'evy case $1<\mu<2$.

Finally, we have studied 
the application to the related Dyson hiearchical spin-glass model
and stressed the similarities and differences.

In the future, we hope to extend the present work to the short-ranged spin-glass model in dimension $d>1$. It is of course clear that the box decimation procedure described in section \ref{sec_box} can only give the exponent (consider boxes of the largest size $b=(L/2)$ as in section \ref{sec_half})
\begin{eqnarray}
\theta^{SR}_{box}(d)= \frac{d-1}{2}
\label{thetasrbox}
\end{eqnarray}
which is known to be a too large upper bound (Eq. \ref{boundtheta}).
For the short-ranged model, it is indeed crucial to have a renormalization procedure that does not fix 
 a priori the boundaries of the correlated clusters independently of the disorder realization. 
For the short-ranged model in $d=1$ (see section \ref{sec_nn1d}), we have already seen that the strong disorder decimation procedure is indeed able to reproduce the correct result $\theta^{SR}_{strong}(d=1)=-1$ below the box value $\theta^{SR}_{box}(d=1)=0 $ of Eq. \ref{thetasrbox}. So we hope that the same phenomenon will occur in dimension $d>1$, and
that an appropriate extension of the strong disorder procedure adapted to the hypercubic lattice will be able to build correlated clusters with boundaries
 adapted to each disorder realization, and to reproduce the correct droplet exponent.

\end{document}